\documentclass[
	a4paper, 
	10pt, 
]{LTJournalArticle}

\usepackage{parskip}
\usepackage{multirow}
\usepackage{float} 

\usepackage{fix-cm}
\usepackage[T1]{fontenc}
\usepackage[fontsize=10.6pt]{fontsize}

\runninghead{Hawley \& Steinmetz, ``Leveraging Neural Representations for Audio Manipulation''} 

\footertext{\textit{Preprint. Accepted as an Express Paper for AES Europe 2023}} 

\title{Leveraging Neural Representations\\ for Audio Manipulation} 

\author{%
	Scott H. Hawley\textsuperscript{1,3}
 \thanks{Corresponding author: \href{mailto:scott.hawley@belmont.edu}{scott.hawley@belmont.edu} \textbf{Preprint:} \today},
         Christian J. Steinmetz\textsuperscript{2}
}

\date{\normalsize\textsuperscript{\textbf{1}}Belmont University, Nashville, TN, USA\\ \textsuperscript{\textbf{2}}Centre for Digital Music, Queen Mary University of London, UK\\ \textsuperscript{\textbf{3}}Harmonai, USA}

\renewcommand{\maketitlehookd}{%
	\begin{abstract}
		\noindent We investigate applying audio manipulations using pretrained neural network-based autoencoders as an alternative to traditional signal processing methods, since the former may provide greater semantic or perceptual organization. 
To establish the potential of this approach, we first establish if representations from these models encode information about manipulations.
We carry out experiments and produce visualizations using  representations from two different pretrained autoencoders. 
Our findings indicate that, while some information about audio manipulations is encoded, this information is both limited and encoded in a non-trivial way. 
This is supported by our attempts to visualize these representations, which demonstrated that trajectories of representations for common manipulations are typically nonlinear and content dependent, even for linear signal manipulations.
As a result, it is not yet clear how these pretrained autoencoders can be used to manipulate audio signals, however, our results indicate this may be due to the lack of disentanglement with respect to common audio manipulations.
	\end{abstract}
}

\begin{document}

\maketitle 


\section{Introduction}

Musical audio production workflows use a variety of parameterized transformations to perform the processing and re-synthesis of audio signals. Examples include the sliders on a multi-band equalizer, dynamic range changes made by adjusting gain or compression, spectral processing, and adjustments made by changing MIDI parameters. The development of increasingly ``intelligent'' music interfaces may be regarded as a pursuit to find 
transformations yielding \emph{representations} that more closely match the perceptual or semantic content of interest to music producers, such as fewer knobs that control high-level aspects of the sound~\cite{stables2014safe}. 

Neural network-based audio autoencoders have shown promise for many applications, including audio coding~\cite{soundstream, encodec} and the transfer of musical style features such as instrument type~\cite{DDSP, RAVE} and audio production details~\cite{deepafx-st}. The type of encoder chosen will produce encoded representations that are typically better suited for some tasks than others. Often such tasks take the form of classification and/or Music Information Retrieval~\cite{HEAR, Jukebox, castellon}. 
We focus on the analysis and synthesis of audio signals by freezing the weights of pretrained autoencoders optimized for audio reconstruction. The latent representations arising in such autoencoders may encode semantic or stylistic information~\cite{castellon}.

\begin{figure}[t!]
\vspace{-0.3cm} 
\begin{center}
\includegraphics[width=0.95\columnwidth]{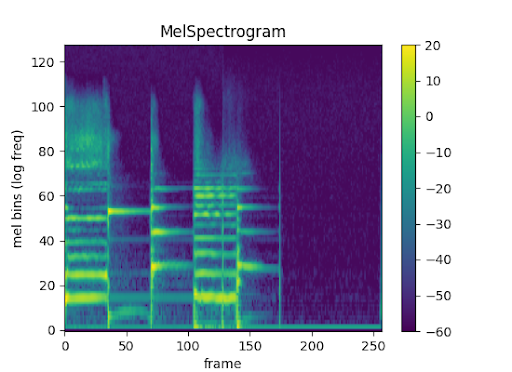}
\end{center}
\vspace{0.1cm} 
\begin{center}\includegraphics[width=0.95\columnwidth]{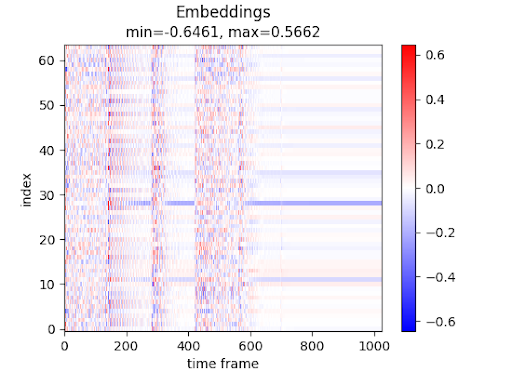}
\caption{Mel spectrogram (top) and 
 latent representation ``spectrogram'' from the DiffAE diffusion autoencoder. 
}
\label{fig:sample_reps}
\end{center}
\end{figure}

To place this work in context, one may consider spectrograms computed via short-time Fourier transforms (STFT), which can be viewed as columns of ``vectors" in a multi-dimensional space. The representations produced by neural network systems can similarly be viewed either as a set of vectors or as columns in "neural [activation] spectrograms" or "feature maps".  Illustrations of such spectrogram-like representations appear in Figure \ref{fig:sample_reps}.

Previous neural network methods for audio effect transformations~\cite{marco,signaltrain,microtcn} have been created in an end-to-end manner using supervised learning. In this paper, we consider transformations {\em only} upon the latent representations of pretrained auto-encoders. By leveraging the encoder and decoder portions for systems that have been pretrained on large datasets, one may be able to discover transformations of latent representations from relatively small datasets.  
In contrast to typical transfer learning approaches that further update the weights of a model, this study wishes to explore the potential of achieving useful signal manipulations by manipulating only the intermediate activations or representations of a frozen pretrained model.

This paper lays the groundwork for the development of few-shot or zero-shot musical audio transformations from a self-supervised training program. Our early attempts at zero-shot style transfer of audio production effects using vector algebra operations on representations from pretrained autoencoders~\footnote{Colab notebook:
\href{https://tinyurl.com/Destructo-ipynb}{https://tinyurl.com/Destructo-ipynb}, Oct. 2022} were largely unsuccessful, motivating the visualizations and classification tests of this paper: we aim for a better understanding of the features of such representation spaces.

We hypothesize that ``good" embedding spaces would also provide for strong coupling to text-based control systems as is seen with text-to-image models~\cite{dalle, stablediffusion,instructpix2pix}. 
Identifying perceptually meaningful transformations in the latent space could also enable the discovery and design of novel audio manipulations, paving the way for new methods of audio effect design and enabling users to more intuitively explore and discover novel sound manipulations~\cite{steinmetz2021steerable}. 

A key challenge for working with latent space representations is to disentangle the different dimensions, e.g., so that user controls such as knobs and sliders have one primary (perceptual or semantic) effect~\cite{HarkonenGANControls, ganspace, RAVE}. Disentangling dimensions has also led to improvements in few-shot voice style transfer~\cite{fewshotvoice}.
 Applications of contrastive methods~\cite{fonseca, MMST_tony} have shown that semantically populating the latent space and disentangling dimensions can be mutually achievable. We choose
 to work with existing autoencoders that have not necessarily 
 been optimized for disentanglement of their latent dimensions, as a way to explore their level of disentanglement.

To investigate this we perform visualization of the representations projected into two and three dimensions using Principal Component Analysis and UMAP~\cite{umap}, as well as conduct classification experiments that aim to quantify both the degree of information encoded about manipulations and how this information is encoded.  
We hope that our investigations will lead to advances in musical audio production that make content-based and semantic operations easier to perform than are currently possible.

\newcommand{\ptt}[1]{\texttt{\textls[-25]{\small{#1}}}}

\begin{table}[t]
\caption{Parameter settings for classification tests. All other settings were defaults except the Compressor used a  ``Ratio'' value of 5.}
\vspace{-0.3cm}
\begin{center}
\fontsize{10.1pt}{10.1pt}\selectfont

\begin{tabular}{l l l c }
\toprule
Abbr & Effect Name & Parameter & Value \\
\midrule
CLN & Clean & -  & - \\
CHS & Chorus & Rate (Hz) & 1  \\ 
CMP & Compressor & Threshold (dB) & -50  \\ 
DLY & Delay & Delay (s) & 0.5  \\ 
DIS & Distortion & Drive (dB) & 25 \\ 
HPF & Highpass Filter & Cutoff (Hz) &  2000 \\ 
LPF & Lowpass Filter & Cutoff (Hz) &  70 \\ 
PS & PitchShift & Semitones & 4  \\ 
RVB & Reverb  & Room Size & 0.8 \\ 
TRV & Time Reverse  & - & - \\
\bottomrule
\end{tabular}
\label{tab:effects_settings_class}
\end{center}
\end{table}

\section{Methods}

\subsection{Models}

The models we studied are two pretrained diffusion autoencoders developed internally by Harmonai~\footnote{These models are not published but are similar to models under development at 
\href{https://github.com/Harmonai-org}{https://github.com/Harmonai-org}}.  
The first model we refer to as ``DiffAE," which uses a 64-dimensional latent representation space; the lower image in Figure \ref{fig:sample_reps} is from the DiffAE model. The second model is a two-stage cascading latent diffusion model~\cite{imagen,zach_stacked} which we refer to as the ``Stacked DiffAE'' or simply ``Stacked'' model. Representations in both stages are 32 dimensional, but those of the second stage are more compact in time, containing a 16$\times$ coarser resolution in time than the first. Unless otherwise noted, we always use the smaller, more compact Stage 2 representations in this paper. 
The (larger) Stage 1 representations used in this paper are obtained on the {\em decoder} side by upsampling the (more compact) Stage 2 representations and performing additional diffusion. 
We have not used any Stage 1 representations from the encoder side.
In addition to the architecture differences, the two models {\em were trained on different datasets}, sampled from an unreleased repository of a variety of music.  In this study it is not our aim to determine which differences in representations are due to particular differences in the pretrained autoencoders, rather we use multiple models to temper any generalizations that might otherwise be drawn from considering only one autoencoder model. 

\begin{table}[t]
\caption{Settings for parameter variation tests, using 32 increments from minimum to maximum, with HPF and LPF values varying logarithmically. All other settings were defaults.}
\begin{center}
\fontsize{10pt}{10pt}\selectfont
\begin{tabular}{lccc}
\toprule
Effect Name & Parameter & Min - Max \\
\midrule
Distortion (DIS) & Drive (dB) & 0 - 30 \\ 
Reverb (RVB) & Room Size & 0.01 - 0.99 \\ 
Highpass Filter (HPF) & Cutoff (Hz) &  50 - 10000 \\ 
Lowpass Filter (LPF) & Cutoff (Hz) &  50 - 10000 \\ 
\bottomrule
\end{tabular}
\label{tab:effects_settings_vary}
\end{center}
\end{table}

\newcommand{\mytt}[1]{\texttt{\textls[-25]{#1}}}

\subsection{Datasets}

We constructed a dataset of 1024 audio samples at 48\,kHz, each 5.4 seconds in length ($2^{18}$ samples).  
We sourced 512 guitar sounds from GuitarSet~\cite{xi2018guitarset}, including  performances of both solo notes and strumming chords. 
The remaining 512 samples were piano sounds sampled from the 2018 subset of the MAESTRO dataset~\cite{hawthorne2018enabling}. 
We chose to use only two musical instruments, solo, so that the results of audio effect manipulations could stand out more clearly than they might if applied to a more widely-varied musical audio dataset. 
All input sounds were mono recordings that were 'doubled' to stereo (because the models expect stereo inputs), and loudness-normalized via \mytt{pyloudnorm}\footnote{\href{https://github.com/csteinmetz1/pyloudnorm}{https://github.com/csteinmetz1/pyloudnorm}}~\cite{steinmetz2021pyloudnorm} before and after passing through audio effects to remove level differences. 
We select nine different audio manipulations using fixed parameters to explore the clustering by effects and four effects for which we investigated the results of varying one key parameter.
All audio effects, except a ``clean'' bypass and a time-reversal, were applied via \mytt{Pedalboard}\footnote{\href{https://github.com/spotify/pedalboard}{https://github.com/spotify/pedalboard}} with settings shown in Table \ref{tab:effects_settings_class} for classification tests and Table \ref{tab:effects_settings_vary} for parameter variation tests.

\begin{figure}[b]
\begin{center}
\includegraphics[bb=90 60 550 310, clip=true, width=0.95\columnwidth]{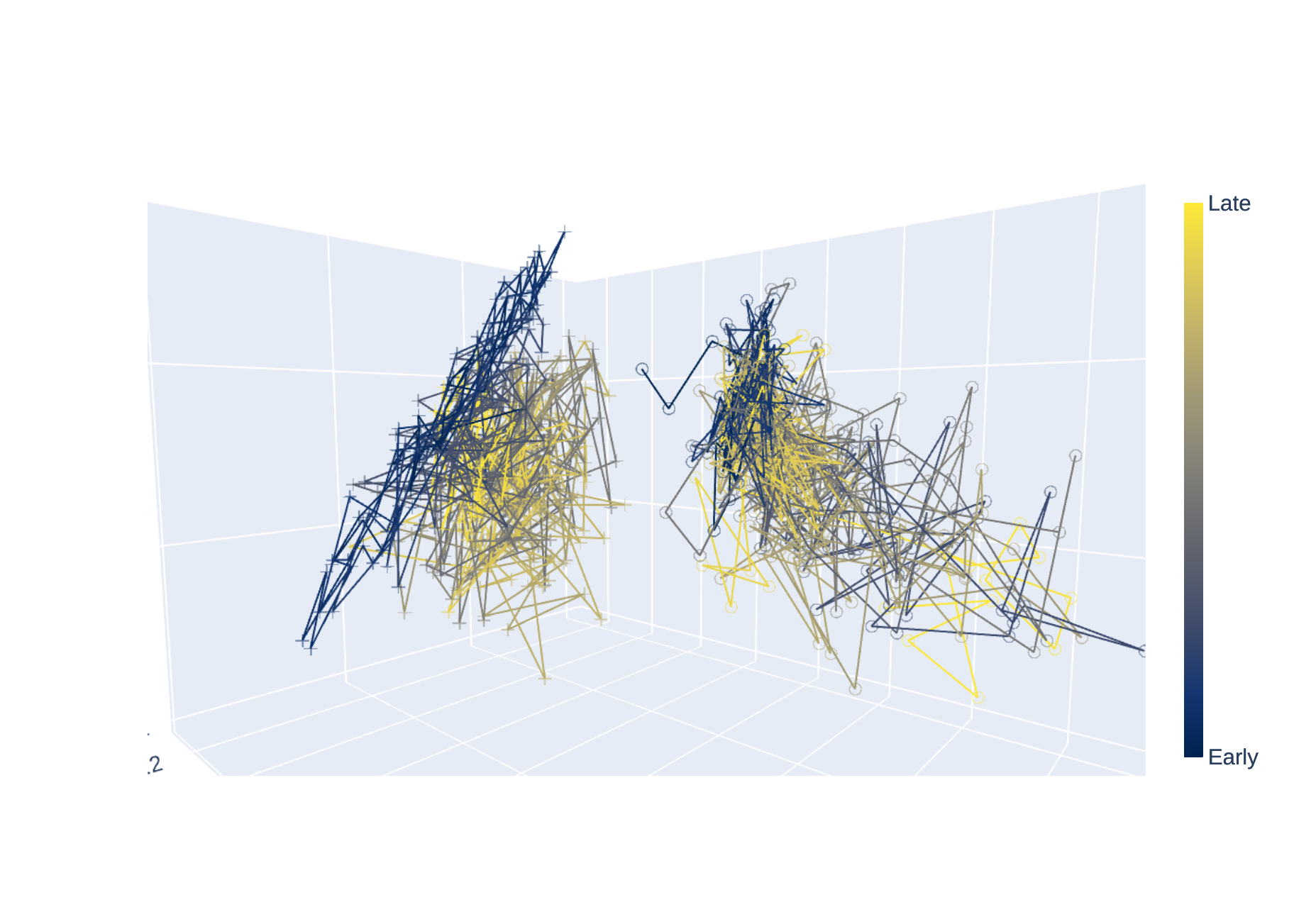}
\caption{PCA plot of time trajectories for one guitar sample (circles) and one piano sample (crosses), for the 32-dimensional Stacked DiffAE model.
}
\label{fig:time_trajectories}
\end{center}
\end{figure}

\begin{figure*}[!ht]
    \includegraphics[width=\textwidth]{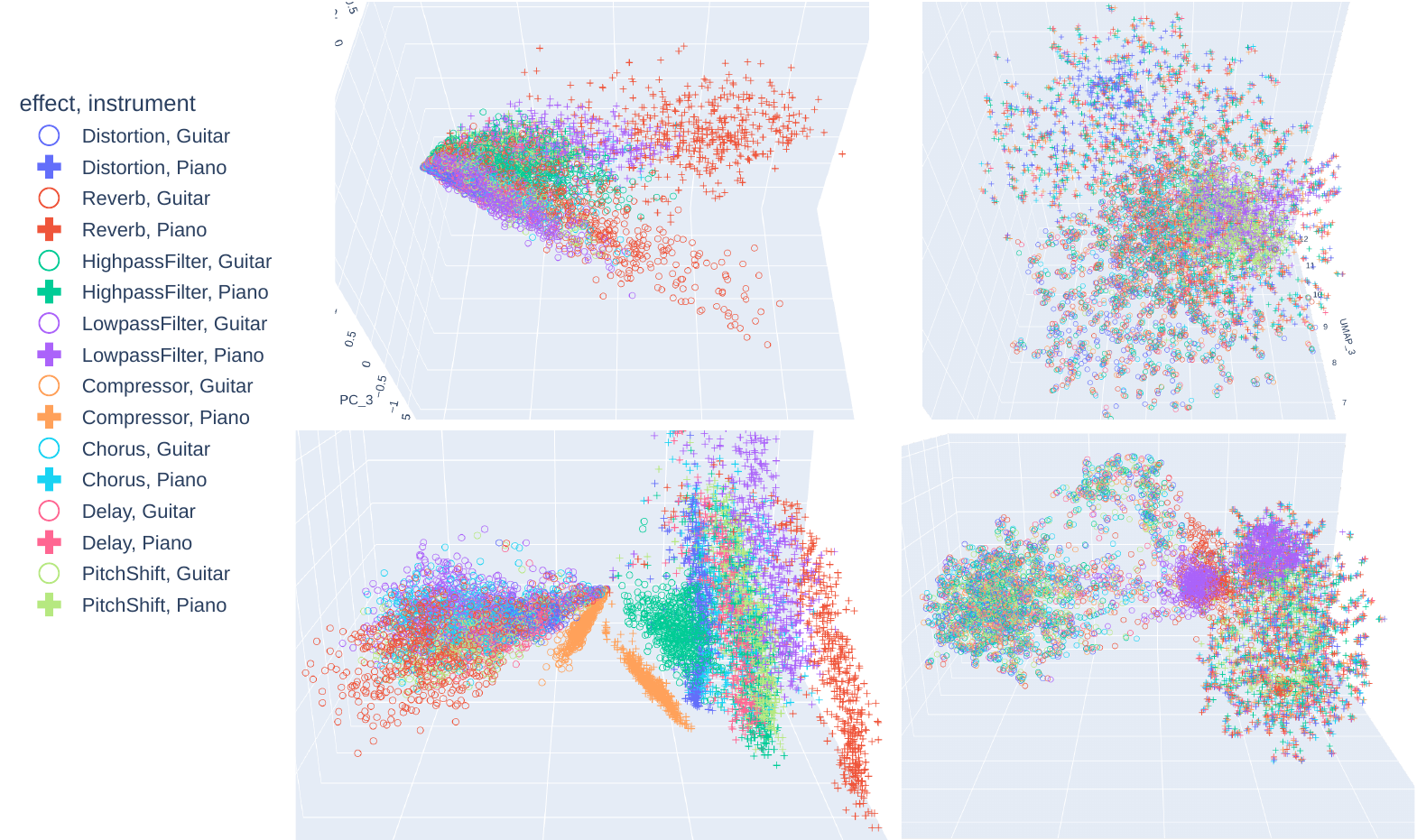}
    \caption[Dummy Caption]{3D projections{\footnotemark} of flattened representations of the DiffAE autoencoder (top row) and stacked model (bottom row),  showing PCA (left column) and UMAP (right column) mappings. In the UMAP plots, one sees that variations due to performance typically are more significant than variations due to audio effects. The two classes of instrument sounds map to largely disjoint areas of the latent space. 
    }
    \label{fig:high-dim}
\end{figure*}

\section{Results}

\subsection{Visualization}
Figure \ref{fig:time_trajectories} shows the trajectories of representation vectors as a function of time,
projected into three dimensions using Principal Component Analysis (PCA). The two trajectories shown are projected using the PCA transformation derived from
the entire dataset. The ``motion'' of these representation vectors describes a complicated path. While methods such as RAVE~\cite{RAVE} employ a ``prior model''
to learn to predict such trajectories autoregressively for music generation, in this study we simply note that the trajectories as seen in 3D PCA projections do not seem to follow any easily intuited pattern.

\begin{figure*}[!ht]
    \includegraphics[width=\textwidth]{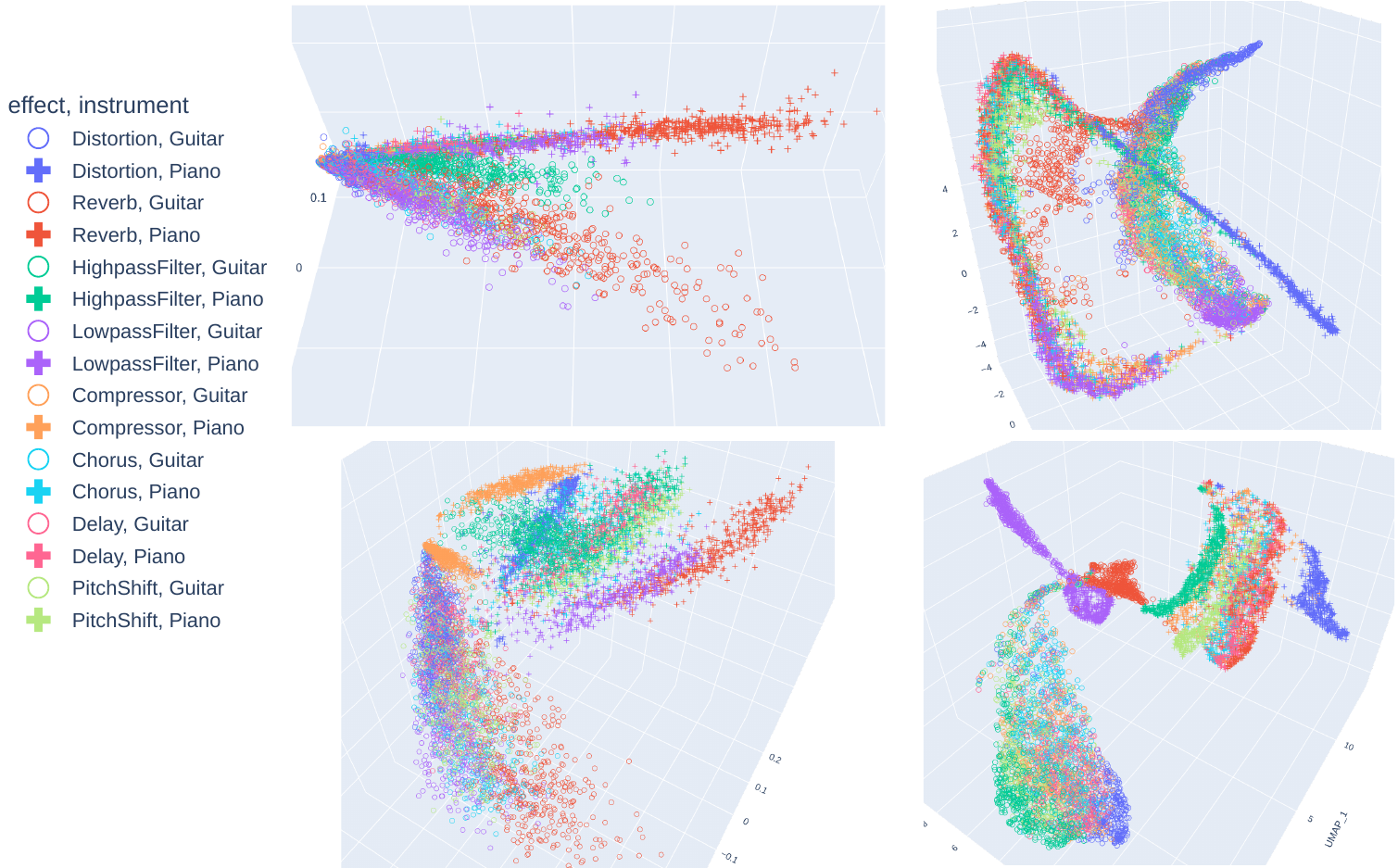}
    \caption{3D projections of time-averaged representations of the DiffAE autoencoder (top row) and stacked model (bottom row), showing PCA (left column) and UMAP (right column) mappings. One sees that the time-averaged representations typically offer better clustering and separation by audio effect than the flattened representations in Figure \ref{fig:high-dim}. As in the previous figure, we observe large-scale clustering according to musical instrument class.}
    \label{fig:time_avg}
\end{figure*}

An alternative view\footnotetext{Interactive versions of 3D plots in this paper are available at \href{https://api.wandb.ai/links/drscotthawley/36hi1day}{https://api.wandb.ai/links/drscotthawley/36hi1day}} results from flattening the representations for all dimensions and times into single vectors that inhabit a high-dimensional space. Figure \ref{fig:high-dim} shows flattened representations colored by audio effect class in PCA and UMAP~\cite{umap} 3D projections. 
The PCA projections in the left column for both the DiffAE (top) and the Stacked (bottom) models show some separation by audio effects class with the latter showing stronger clustering by audio effects. However, the UMAP images in the right column of Figure \ref{fig:high-dim} show that except for one or two effects, the clustering of flattened representations is not class-determined: the variation due to musical 
performance (of the guitar or piano piece) plays a greater role in the data representations, with audio effects being a small perturbation. 

To obtain a stronger ``style'' signal to investigate the separation by audio effect class, we time-average the representations and plot them in Figure \ref{fig:time_avg}. The resulting PCA data (left column) appears similar to the previous Figure, however, the UMAP plots (right column) now show much stronger grouping by audio effects than in the previous figure. This suggests that time averages of the representations may serve as a useful proxy for ``style'' information, and at the very least that the representations of these autoencoder models do encode some ``semantic'' 
structure with respect to audio effect class. 

\begin{figure}[t]
\begin{center}
\includegraphics[bb=150 00 530 320, clip=true, width=0.9\columnwidth]{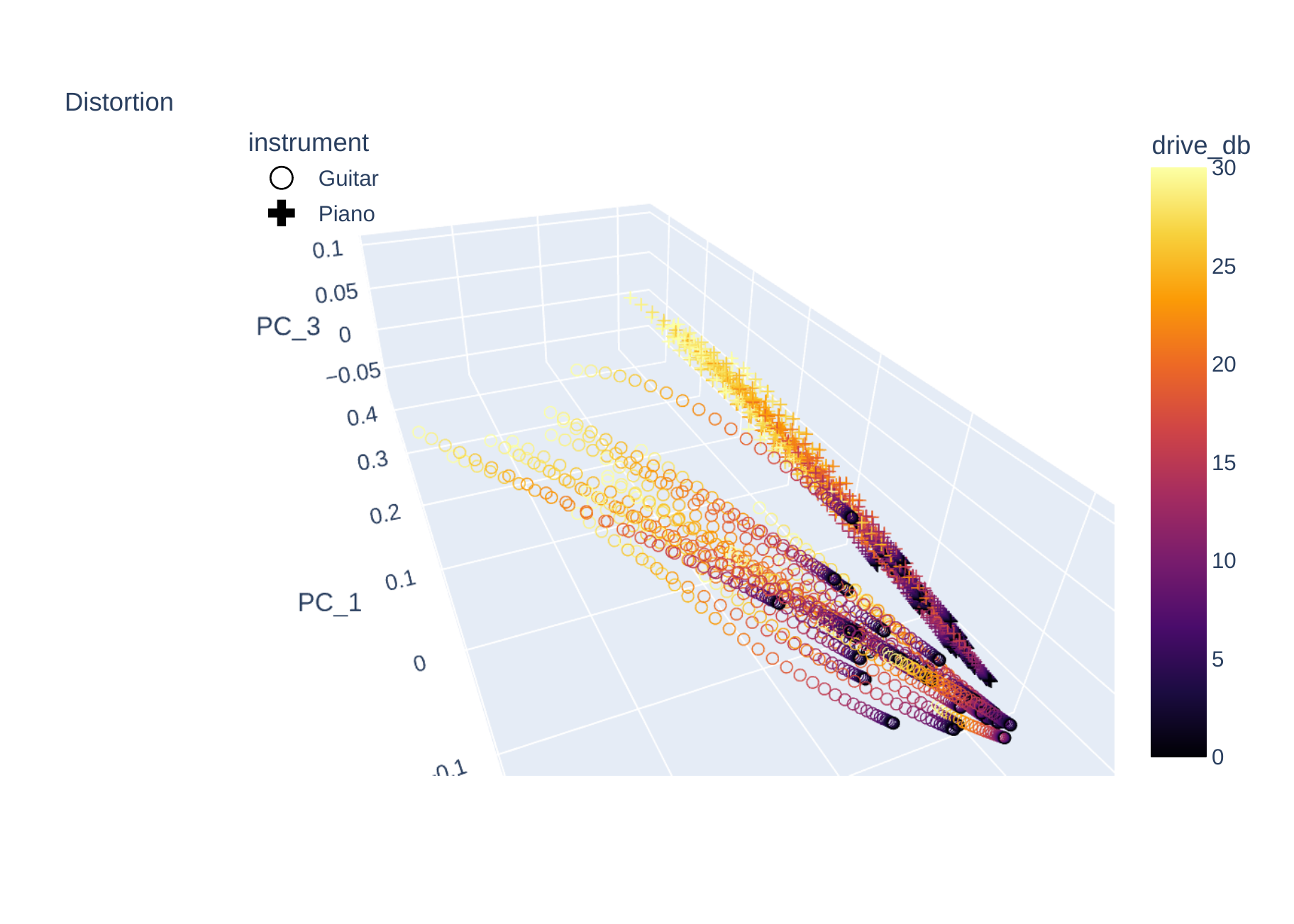} \\
\vspace{-.6cm}
\includegraphics[bb=150 60 530 325, clip=true, width=0.9\columnwidth]{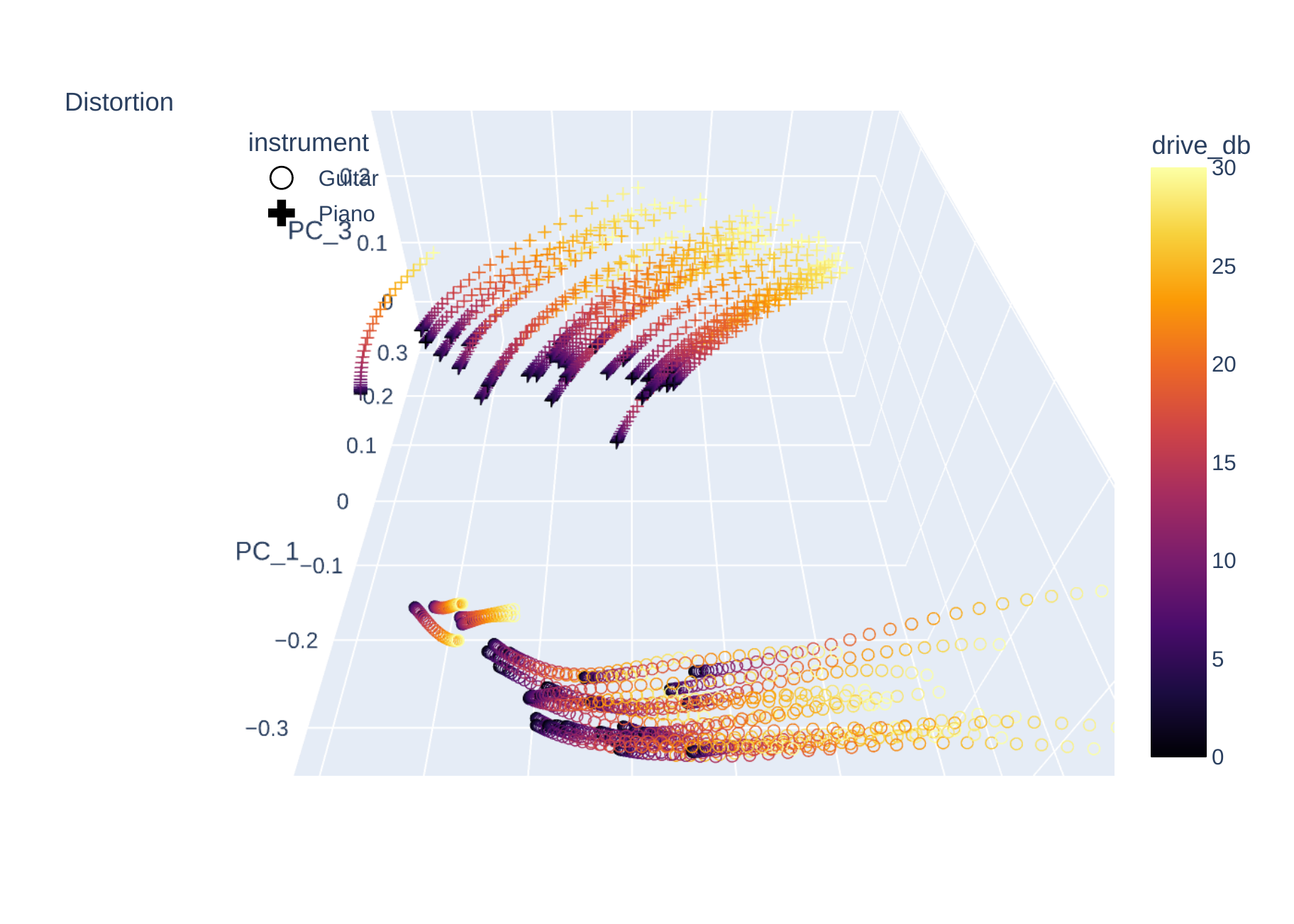}
\vspace{0.3cm}
\caption{Varying the Distortion effect on 16 guitar and 16 piano samples, for the DiffAE (top) and Stacked DiffAE (bottom) encoders. Here we show PCA plots of time-averaged embeddings. 
}
\label{fig:distortion_knob_trajectories}
\end{center}
\end{figure}

\begin{figure}[t]
\begin{center}
\includegraphics[bb=60 60 510 330, clip=true,width=0.95\columnwidth]{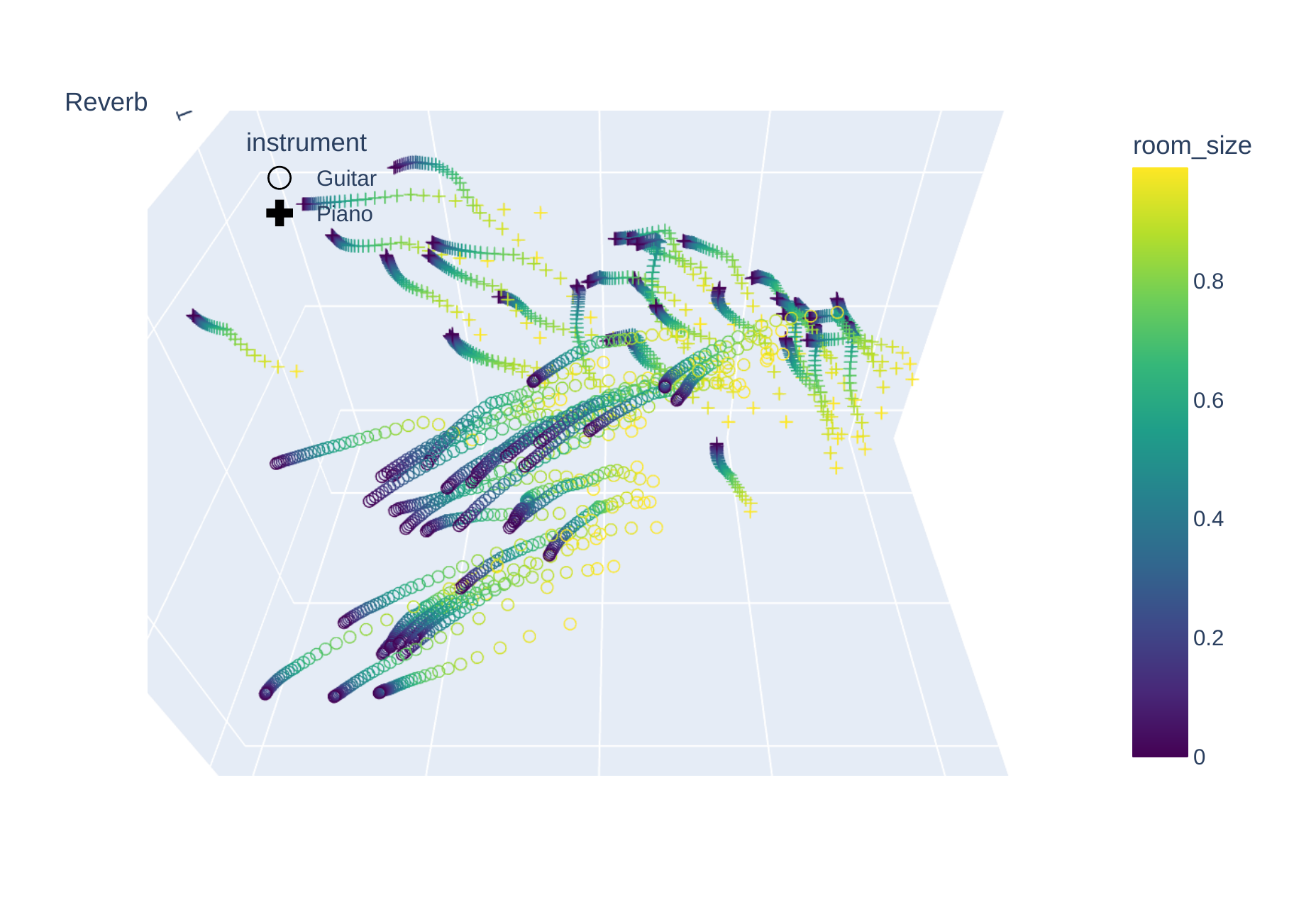}
\vspace{0.3cm}
\caption{PCA plot of parameter adjustment paths for Reverb effect using time-averaged representations from the Stacked model, for guitar (circles) and piano (crosses).}
\label{fig:reverb_knob_trajectories}
\end{center}
\end{figure}

\begin{figure}[b]
\vspace{0cm}
\begin{center}
\includegraphics[bb=60 60 510 330, clip=true, width=0.9\columnwidth]{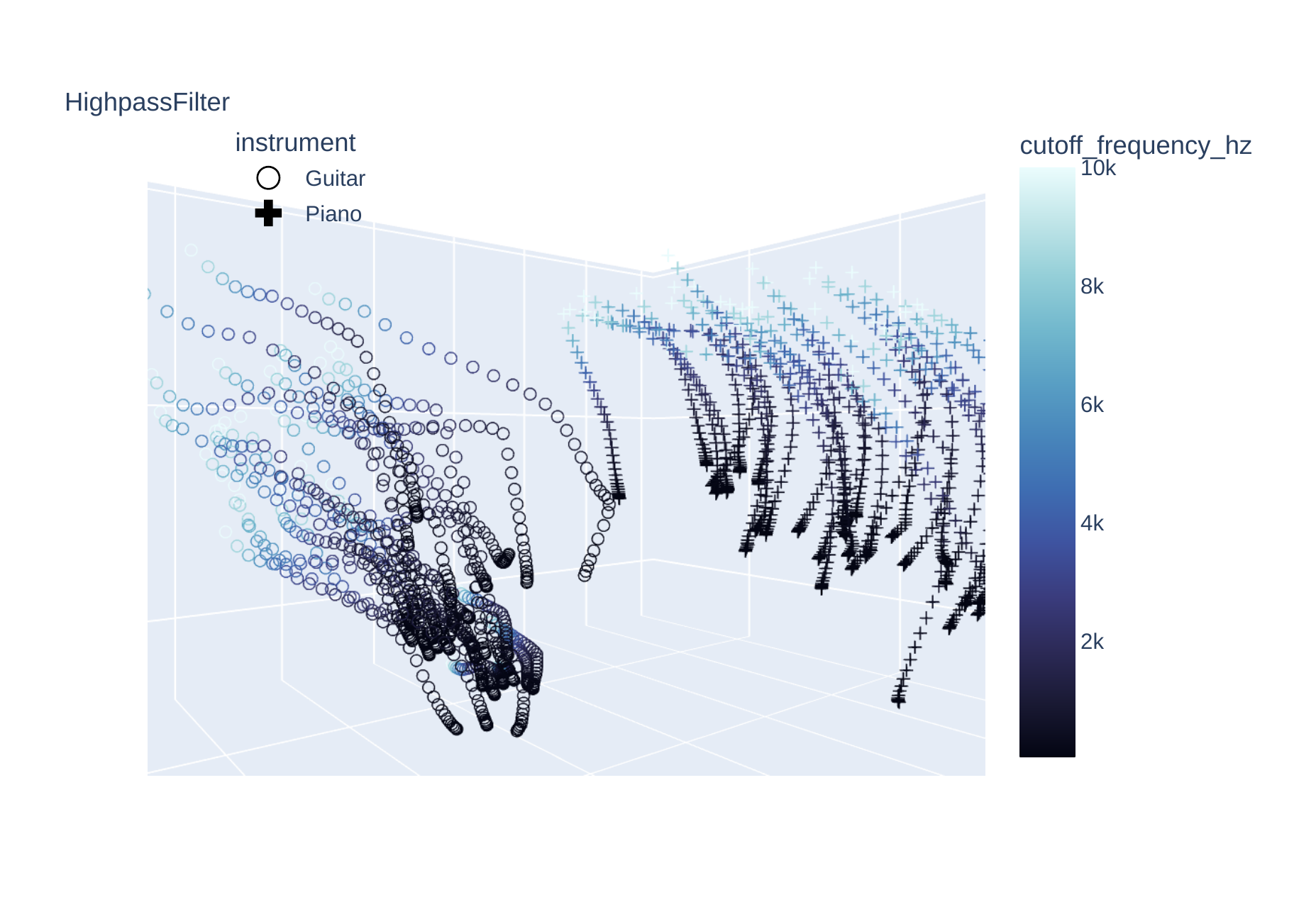}\\
\vspace{0.1cm}
\includegraphics[bb=60 60 510 330, clip=true, width=0.9\columnwidth]{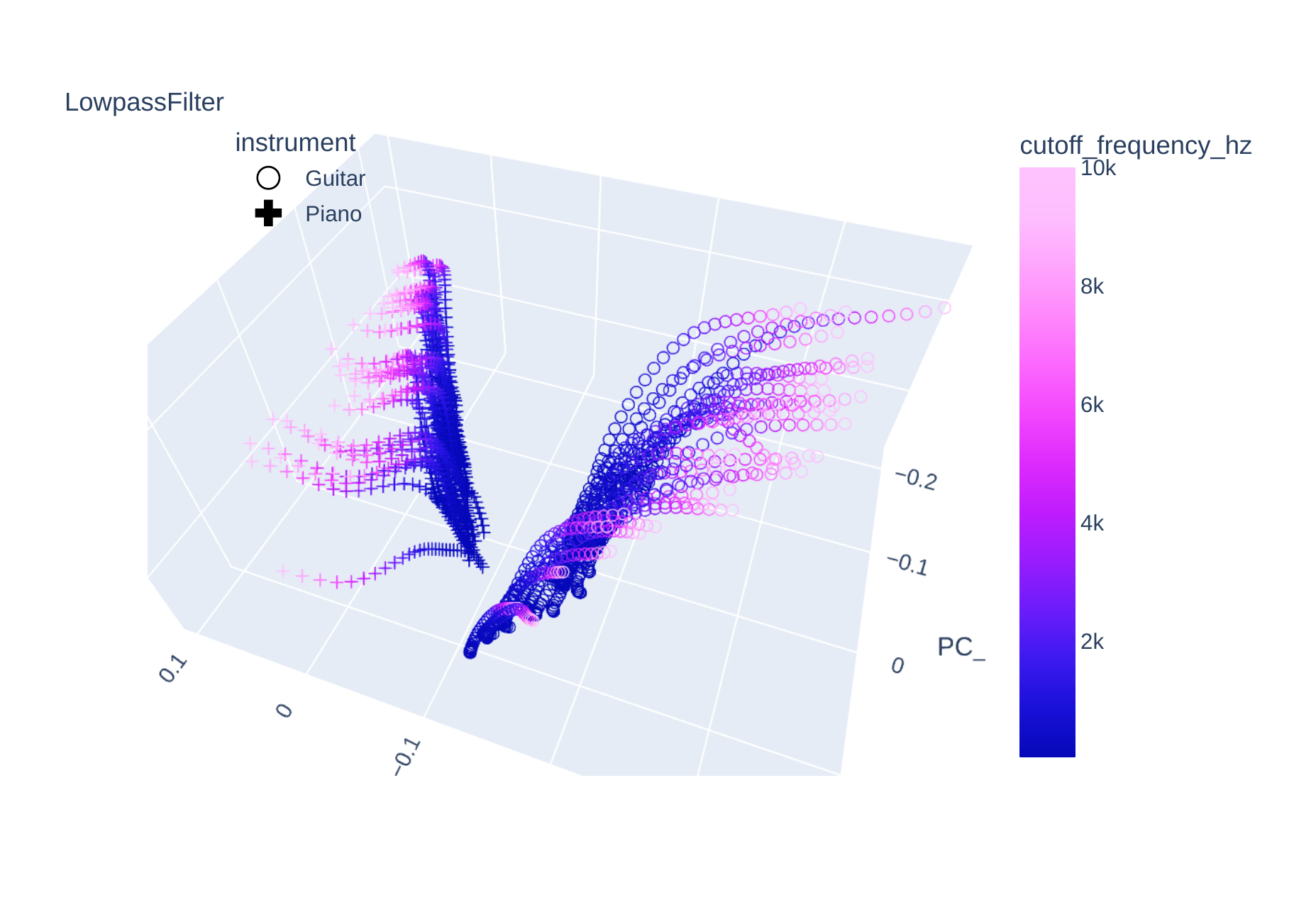}
\vspace{0.3cm}
\caption{PCA plots of time-averaged representations in the Stacked model,  adjusting Highpass (top) and Lowpass (bottom) filters for guitar (circles) and piano (crosses)}
\label{fig:lpf_knob_trajectories}
\end{center}
\vspace{0cm}
\end{figure}

\begin{figure}[b]
\begin{center}
\includegraphics[bb=00 00 750 400, clip=true,width=1.1\columnwidth]{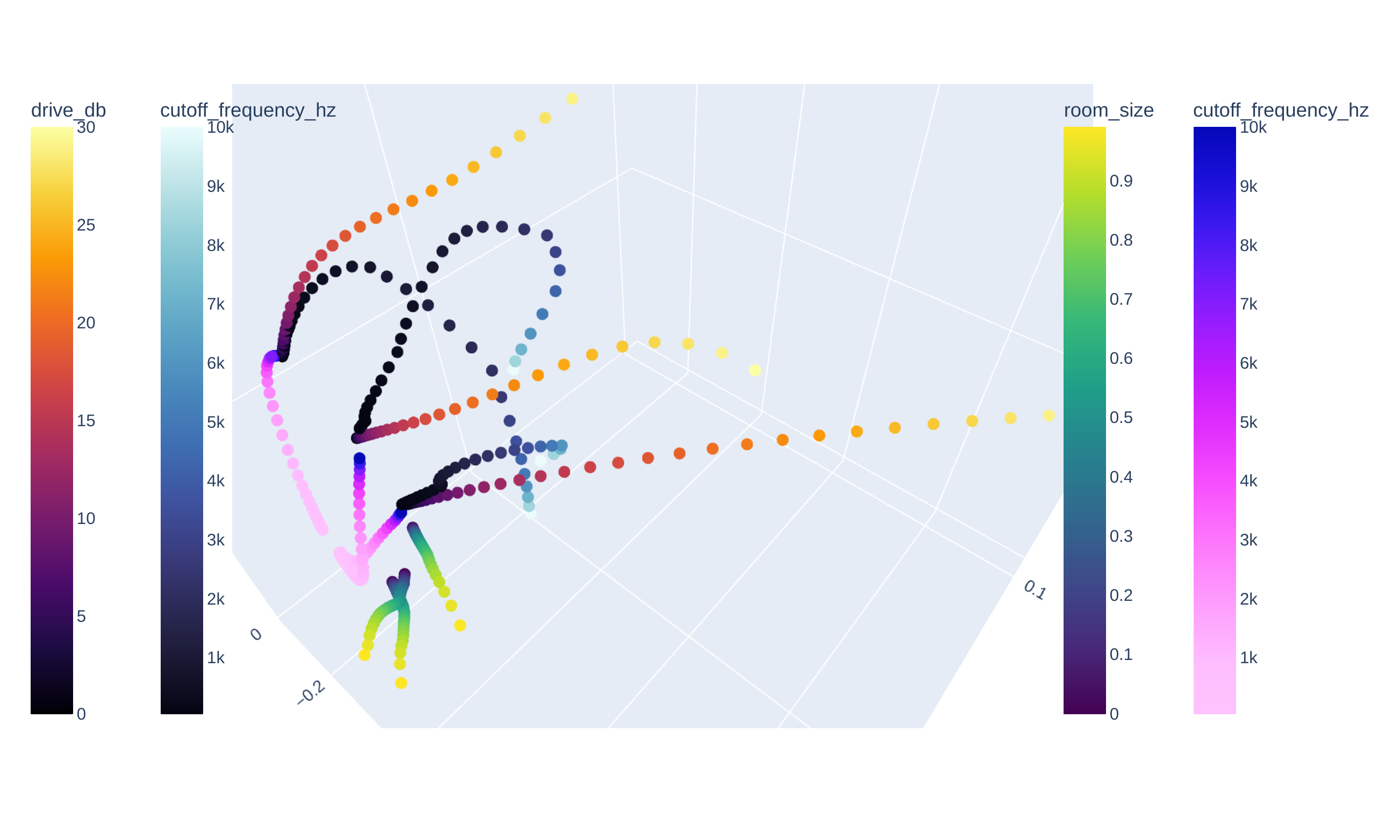}
\includegraphics[bb=00 60 750 400, clip=true, width=1.1\columnwidth]{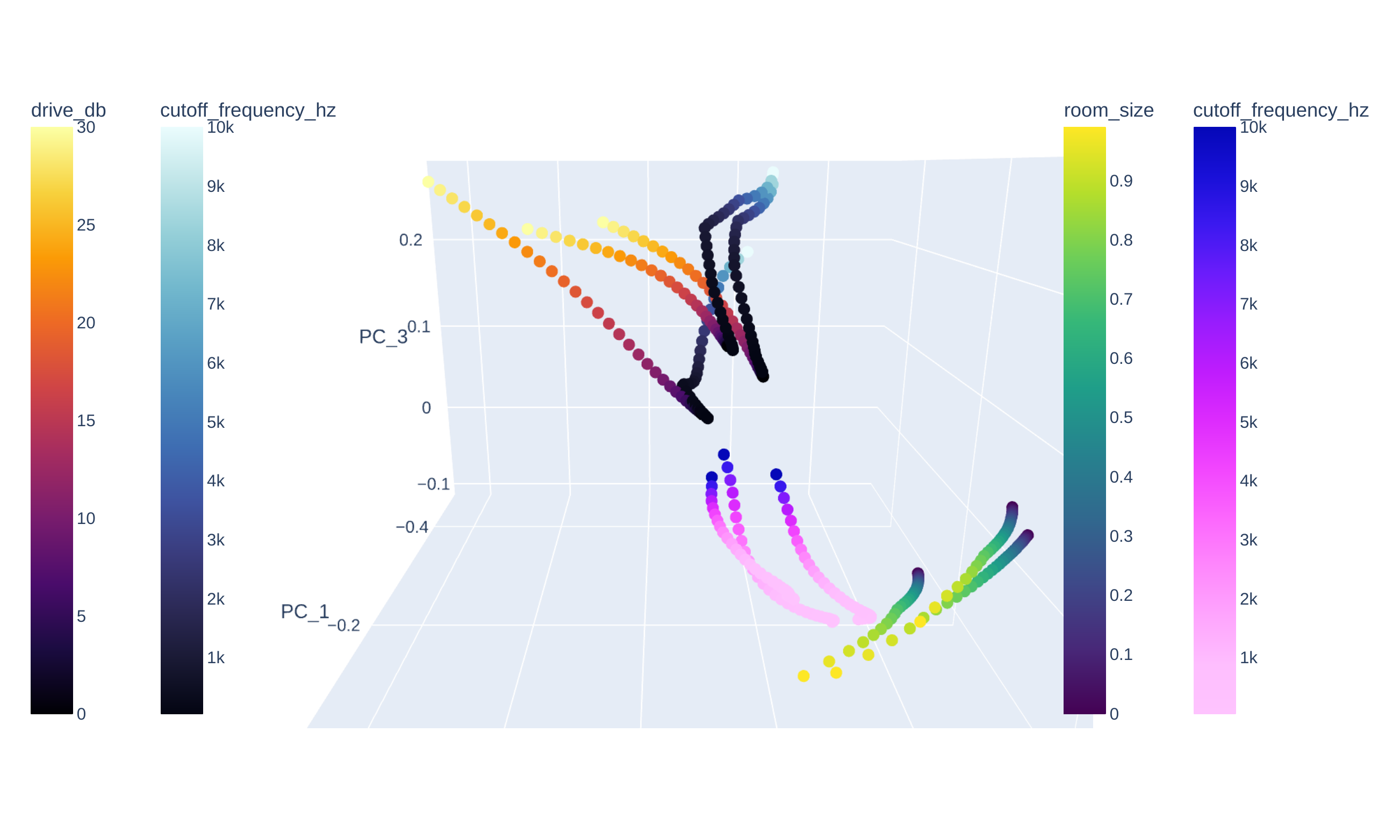}
\vspace{0.2cm}
\caption{Multi-effects trajectories for DiffAE (top) and Stacked DiffAE (bottom) models.
}
\label{fig:multi_effects_trajectories}
\end{center}
\end{figure}

A final visualization study explores the variation in embeddings due to effect parameter values. Figure \ref{fig:distortion_knob_trajectories} through \ref{fig:lpf_knob_trajectories} show the effects of varying key ``knob'' values for Distortion, Reverb, Highpass and Lowpass Filter effects. A primary observation is that varying an affect parameter typically results in a nonlinear path through the space, even for ``linear'' effects such as the filters. 
This helps explain the unsuccessful results of earlier attempts at few-shot neural audio style transfer based on algebraic operations on representation vectors.
Figure \ref{fig:multi_effects_trajectories} combines the display of multiple effects parameters applied to a few audio samples. Again, one sees curved paths
even for ``linear'' effects such as HPF, whereas perhaps unexpectedly the Distortion paths tend to appear comparatively unidirectional.

\subsection{Manipulation classification}

To quantitatively measure the level of information about signal manipulations captured by our autoencoders we use manipulation classification as a proxy task.
Similar to the evaluation methodology employed in self-supervised representation learning, we use a linear probe on top of the embeddings from frozen pretrained encoders to perform classification~\cite{chen2021empirical}. 
In our setup, we train linear probes on the 10-way classification task, predicting which of the 9 different audio manipulations was applied to the source audio or if no manipulation is present. 
Linear probes consist of a single layer that maps from the embedding space to the number of classes. This results in a parameter count proportional to the embedding dimensionality.

Since our encoders produce a sequence of representations (in time), we adopt two different approaches to feeding the extracted embeddings to the linear probe. 
First, we consider the time average of all embeddings, which has the effect of reducing variance across time frames. 
However, time averaging may remove information critical for detecting audio effects if the effect exhibits behavior over larger time scales, for example, reverberation. 
To account for this case, we also consider the complete flattened sequence of embeddings.

\setlength{\tabcolsep}{4.5pt}
\begin{table*}[!ht]
    \centering
    \begin{tabular}{l r r c c c c c c c c c c c} \toprule
        
        \multirow{2}{*}{Encoder}  & \multirow{2}{*}{Dim}  & \multirow{2}{*}{Probe} & \multicolumn{10}{c}{Class-wise Accuracy (\%)}\\ \cmidrule(lr){4-13} 
                 &           &                       & CHS      & CLN      & CMP      & DLY      & DST      & HPF      & LPF      & PS       & RVB      & TRV      & AVG\\ \midrule
                        VGGish-T  & 128   &  1.3\,k  & 98.0     & 80.6     & 90.1     & 81.3     & 91.4     & 94.5     & 96.9     & 86.7     & 90.4     & 94.5     & 92.9 \\
                        VGGish-F  & 2176  & 21.8\,k  & 98.0     & 76.5     & 85.2     & 76.8     & 92.1     & 90.1     & 96.9     & 83.3     & 88.7     & 95.1     & 90.3 \\ \midrule
                        Stacked2-T &  32   &  0.3\,k  & 24.0     & 54.9     & 28.1     & 34.9     & 86.6     & 78.5     & 95.0     & 66.7     & 79.3     & 32.0     & 57.7 \\
                        Stacked2-F & 16384 & 163.9\,k & 45.2     & 42.2     & 30.2     & 65.1     & 84.7     & 80.9     & 94.4     & 68.2     & 71.3     & 39.2     & 63.1 \\  \midrule
                        Stacked1-T & 32   &  0.3\,k  & 29.9     & 59.2     & 27.1     & 26.5     & 88.6     & 71.4    & 96.9      & 54.2     & 74.8     & 36.8     & 57.4 \\
                        Stacked1-F &262144&2621.5\,k & 14.6     & 37.6     & \phantom{0}9.4     & 29.2               &  41.9    & 59.1     & 88.7     & 24.5     & 58.5     & 48.6     & 41.1 \\ \midrule
                        DiffAE-T &  64   & 0.7\,k   & 11.2     & 32.5     & \phantom{0}3.6     & \phantom{0}9.3     & 92.9     & 46.1     & 86.9     & 31.0     & 63.5     & 27.1     & 40.3 \\
                        DiffAE-F & 131072& 1310.7\,k& \phantom{0}4.4     & 32.7     & 47.7     & 16.3     & 73.6     & 17.0     & 27.0     & 19.3     & 18.6     & 25.6    & 28.9 \\ 
        \bottomrule
    \end{tabular}
    \caption{Effect classification accuracy when training a linear probe on top of normalized representations from pretrained encoders. T denotes time-averaged embeddings and F denotes flattened embeddings. Best-performing models are denoted in boldface.}
    \label{tab:classifier_results}
\end{table*}

In our evaluation, we compare representations from the Stacked autoencoder (Stage 1 and Stage 2), as well as the DiffAE. As non-autoencoding baseline we also use embeddings extracted from the pretrained VGGish model\footnote{\href{https://github.com/hearbenchmark/hear-baseline}{https://github.com/hearbenchmark/hear-baseline}}~\cite{hershey2017cnn}.
Similar to the autoencoders, the VGGish was not trained to explicitly capture information about audio manipulations. However, unlike the autoencoders, it does not enable reconstruction of the original signal given an encoded sequence.

All probes are trained using a batch size of 32 with the AdamW optimizer and a learning rate of $3 \cdot 10^{-4}$ for a maximum of 500 epochs. 
We use early stopping with a patience of 50 epochs monitoring the validation accuracy. 
Results from these classification tests are shown in Table \ref{tab:classifier_results}.  
While all features enable classification accuracy higher than random guessing, we find that the VGGish features significantly outperform all of the autoencoder representations. 
This indicates that all representations encode some information that is predictive of audio manipulations, however, there is comparatively less information captured in the autoencoder representations when used both in the time-averaged and flattened configuration.

\subsection{Dimension masking}

While the multi-class classification experiment provides some insight into the level of information about audio manipulations, it does not provide any insight into \emph{how} this information is encoded. 
Developing an understanding of how information is encoded is critical for leveraging the embedding space for signal manipulation.
To approach this question, we designed a second classification experiment. 

In this experiment, we consider a binary classification task where a linear model is trained to classify if a given effect is present or if the signal is clean. 
To understand how information is encoded within the pretrained representation, we mask one dimension of the representation and then train a classifier. 
We then repeat this process for each dimension of the representation and for each of the 9 effects, measuring any decrease in classification accuracy. If masking one dimension significantly decreases performance, it indicates that information predictive of the manipulation in question is encoded in this dimension. 

Due to its superior performance in the previous experiment, we consider the Stacked autoencoder model, which features a 32-dimensional representation. 
Therefore we train 32 linear classifiers for each of the 9 effects, masking one dimension of the representation each time to produce a total of 228 models. 
Similar to the previous experiments, we use a batch size of 32 along with the AdamW optimizer and a learning rate of $3 \cdot 10^{-4}$. 
All models are trained for 100 epochs and we select the best model for evaluation based on the validation accuracy.

\begin{figure*}
    \centering
    \includegraphics[width=\linewidth,trim={0.2cm 0.9cm 0.2cm 1.6cm},clip]{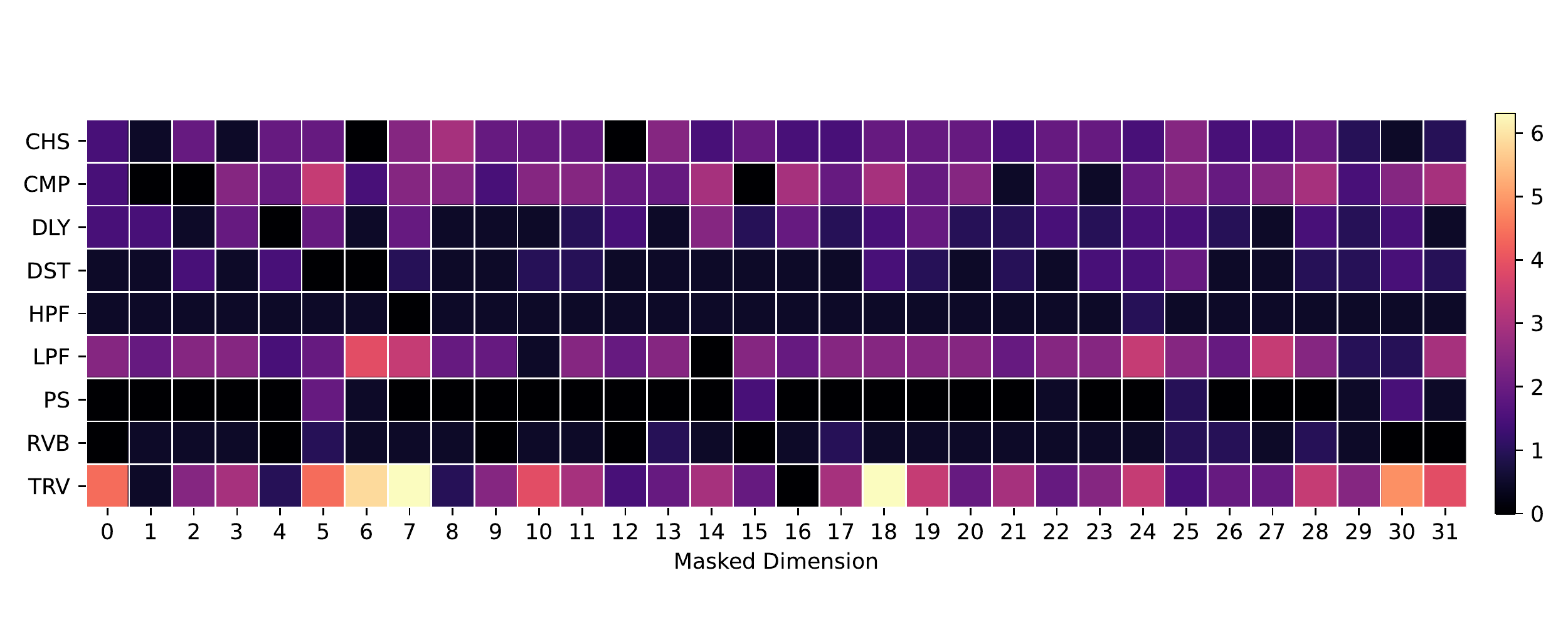}
    \caption{Evaluation of the relative importance of each dimension of the Stacked autoencoder (Stage 2) ($D=32$) using time-averaged embeddings when detecting the presence of each audio effect. The colormap indicates the decrease in the number of percentage points for the binary classification accuracy when masking one dimension of the representation as compared to the best-performing configuration for each effect type.}
    \label{fig:dim-test}
\end{figure*}

We report the results in Figure~\ref{fig:dim-test}, where the color map indicates the number of percentage points lost by masking the associated dimension as compared to the best-performing model for that effect.
We note that for all dimensions and effects, it appears that no single dimension is particularly predictive. 
In the worst case, masking dimension 7 and 18 decreased the classification accuracy for time reversal by $\approx6$ percentage points.
However, it seems that masking any dimension of the representation caused a larger decrease in accuracy for certain effects, such as time reverse, lowpass filter, chorus, and compressor, while effects like highpass filter, pitch shift, and reverberation saw little variation. 
From these results we conclude that disentanglement of the representation with respect to the manipulations studied is likely low.

\section{Summary}
We explored the latent spaces of two pretrained audio autoencoders by manipulating musical audio samples via multiple classes of audio effects. The inner representations of these models show some clustering according to audio effects when one considers the time average of the effects, whereas the full representations in flattened form tend to cluster more strongly based on each individual musical performance. In all cases, the space tended to be divided by musical instrument type (e.g. guitar samples on one side, piano on another). 

To provide a quantitative measurement of the separation by audio effect, we applied classification tests using a linear probe, comparing against a VGGish model as a baseline. We found that the frozen autoencoder representations tended to perform significantly worse than the VGGish model for classification by audio effects, with the time-averaged representations performing better than the flattened ones -- in agreement with our observations of the visualizations. 

By varying the parameters of audio effects, we observe that the resulting path of time-averaged representations through latent space tends to be nonlinear, even for ``linear'' effects such as Highpass and Lowpass filters. Thus representation-based methods for audio manipulation must take this inherent nonlinearity into account.

The findings of this paper may inform future efforts to develop efficient, sophisticated methods for musical audio production which can take advantage of the tendency for neural network autoencoders to encode semantic content. We speculate that additional work in disentangling the dimension of the latent spaces may yield improved results for audio production workflows. 

\section{Acknowledgments}
This research was supported by Stability AI via Harmonai. We wish to thank Zach Evans for use of his pretrained autoencoders. C.S. is funded in part by UKRI and EPSRC as part of the ``UKRI CDT in Artificial Intelligence and Music,'' under grant EP/S022694/1. S.H. acknowledges Max Ortner for helpful discussions.

\clearpage 

\printbibliography 

\end{document}